\documentclass[11pt]{article}
\usepackage{amsmath}
\usepackage{epsf,afterpage,epsfig}

\begin{document}
\begin{flushright}
BI-TH-97/51 \\
NBI-TH-97-55\\
April 1998
\end{flushright}

\begin{center}
\vspace{24pt}

{\LARGE \bf The Ising model on a Quenched Ensemble \\
  \vspace{5pt} of {\boldmath $c=-5$} Gravity Graphs}
\vspace{24pt}

{\Large \sl K.~N.~Anagnostopoulos$\,^a$, P.~Bialas$\,^b$ 
            \,{\rm  and}\, G.~Thorleifsson$\,^c$}  

\vspace{12pt}
\begin{center}
\begin{tabular}{rl}
$^{(a)}$ & The Niels Bohr Institute, Blegdamsvej 17, DK-2200 Copenhagen,
           Denmark, \\
$^{(b)}$ & Institute~of~Computer~Science, Jagellonian University, \\  
         &  Nawojki 11, 30-072 Krakow, Poland, \\
$^{(c)}$ & Fakult\"{a}t f\"{u}r Physik, Universit\"{a}t Bielefeld,
           D--33615 Bielefeld, Germany. \\
\end{tabular}
\end{center}

\vspace{12pt}

\begin{abstract} 
We study with Monte Carlo methods an ensemble of $c=-5$ gravity
graphs, generated by coupling a conformal field theory with central
charge $c=-5$ to two-dimensional quantum gravity.  We measure the
fractal properties of the ensemble, such  as the string susceptibility
exponent $\gamma_s$ and the intrinsic fractal dimensions $d_H$.  We
find $\gamma_s = -1.5(1)$ and $d_H = 3.36(4)$, in 
reasonable agreement  with theoretical predictions.  In addition, we
study the critical behavior of an Ising model on a {\it quenched}
ensemble of the \mbox{$c=-5$} graphs and show that it agrees, within
numerical accuracy, with theoretical predictions 
for the critical behavior of an Ising model coupled {\it
dynamically} to two-dimensional quantum gravity, provided the total central
charge of the matter sector is $c=-5$.  From this we conjecture that
the critical behavior of the Ising model is determined solely by the
average fractal properties of the graphs, the 
coupling to the
geometry not playing an important role.  
\end{abstract}
\end{center}

\vspace{15pt}


\section{Introduction}

Randomness in statistical systems arises in a variety 
of situations and is a very rich and complex subject.
Quenched randomness is frequently used in studying the role of 
impurities and inhomogeneities in real physical systems
where the characteristic time-scale of the disorder is much
longer than other dynamics of the system.  Annealed 
randomness, on the other hand, arises naturally in
studies of fluctuating geometries, such as 
two-dimensional quantum gravity or fluid membranes,
where the disorder is dynamically modified by interaction
between the geometry and matter fields living on the surfaces. 

For a statistical system coupled to random disorder,
either in a quenched or annealed approach, the main
question is to assess the effect randomness 
has on the critical behavior of the pure system.  One 
prediction in this direction is the Harris conjecture
\cite{harris} which states that randomness changes the
values of critical exponents only if the specific heat
exponent $\alpha$ of the pure system is positive.  
This conjecture has been studied 
in many models with quenched disorder,
such as the $2d$ Ising model \cite{dots} (where the
Harris criterion is ambiguous as $\alpha = 0$) and the
Potts model \cite{bondpott}. For both models a change in the 
critical behavior is observed.  

All the above mentioned studies deal with weak disorder.
More recently the critical behavior of systems on
lattices with fractal structure very different from
a flat surface has been investigated.  Such systems
arise naturally when matter, in the form of
conformal field theories, is coupled 
to two-dimensional quantum gravity.  These models can be 
studied either in a continuum formulation, by Liouville field
theory, or using discretized approaches like, for example,
models of dynamical triangulations, formulated either 
as matrix models or studied with numerical simulations.
For these systems the disorder is, however, different from the
one discussed above in that it is annealed, i.e.\ the models
couple dynamically to fluctuations in the geometry.

A remarkable degree of universality does emerge for models
coupled to two-dimensional quantum gravity. 
Namely, the change in the critical behavior
of the systems, and their back-reaction on the geometry, 
only depends on the total central charge
of the matter sector.  This manifests itself in the so-called
KPZ scaling relation which describe how 
the scaling dimensions of conformal operators 
are changed by the interaction with
gravity \cite{kpz}.  Moreover, this universality also extents to
the fractal structure of the surfaces, 
from which we derive the string susceptibility exponent 
$\gamma_s$ and the fractal dimension $d_H$. 

In view of this universality it is tempting to conjecture that
the critical behavior of a particular system, 
when coupled to a fluctuating geometry,
only depends on the (average) fractal structure of the surface.
Details of the interaction between the system and the geometry, 
or the geometrical fluctuations, are not important as such ---
they only serve the purpose of defining the average fractal
geometry.  If this conjecture is true it implies 
that how the average over disorder is performed, i.e.\
that the disorder is annealed, is not essential.  In particular,
predictions of the KPZ scaling relation for the change in the
critical behavior should just as well apply to  models
with quenched disorder, {\it provided the quenched average
is taken over the same ensemble of disorder as is generated
in the annealed approach}.

There are some recent simulations that have 
addressed the question of the critical behavior of
spin models on a quenched ensemble of graphs 
generated by two-dimensional quantum gravity.
Both the Ising model \cite{bhj} and the 10-state Potts model
\cite{bjj} have been studied on an ensemble of pure gravity 
graphs ($c=0$).  For the Ising model a critical behavior
compatible with an Ising model coupled dynamically to
gravity was found, although the accuracy of the results
is not sufficient to rule out the conjecture discussed above. 

The goal of this paper is two-fold.  First, we want to investigate
the fractal geometry of two-dimensional quantum gravity coupled to a
conformal field theory with central charge $c=-5$.  More precisely, we
want to determine the fractal dimension of the corresponding surfaces,
using recently developed finite-size scaling methods
\cite{hausd,janhaus} and to compare it to the (contradictory)
theoretical predictions that exist \cite{anhaus1,anhaus2}.  Second, we
want to investigate the critical behavior of an Ising model on a
quenched ensemble of $c=-5$ graphs and to compare it with
predictions from Liouville theory, for the critical behavior of an
Ising model coupled dynamically to two-dimensional quantum gravity,
and to verify, or disprove, our conjecture about the effect of the
disorder.  Our motivation for choosing $c=-5$ is that both its
predicted fractal structure and the critical
behavior of the Ising model is substantially different from both 
a flat space and for a pure two-dimensional quantum gravity.  This
makes these different critical behavior easier to distinguish in
numerical simulations.

The paper is organized as follows: In Section~2 we study the 
fractal properties of a $c=-5$ conformal field theory coupled
to two-dimensional gravity.  We define the model in Section~2.1 
and discuss the details of the simulations in Section~2.2.
In Sections~2.3 and 2.4 we present our measurements of
the string susceptibility exponent $\gamma_s$ and of the 
fractal dimension $d_H$.  And in Section~2.5 we comment on 
how this particular ensemble of graphs differs from other
types of graphs frequently used in studying disordered system. 
The second part of the paper deals with an Ising model on the
$c=-5$ graphs in a quenched approach.   
In Section~3.1 we discuss the prediction
from Liouville theory for the critical behavior of an Ising
model coupled dynamically to two-dimensional gravity.  
In Section~3.2 we discuss details of the simulations and 
the observables we use to probe the critical behavior.  
In Sections~3.3 and 3.4 we determine the critical temperature
of the Ising model and the corresponding critical exponents.  
Finally, in Section~4 we summarize and discuss our results.


\section{Geometrical properties of {\boldmath $c=-5$} graphs}

\subsection{Dynamical triangulations coupled to scalar matter fields}

The model we study, and use to define our ensemble of random
surfaces, is a discretization of a free bosonic string theory
embedded in $D$--dimensions (for an excellent review see e.g.\
Ref.~\cite{review}).  In the continuum formulation
(Liouville theory) the partition function of two-dimensional quantum
gravity coupled to a conformal field theory, with 
central charge $c$, is defined as 
\begin{equation}
Z(\bar{\mu}) \; = \; \int {\cal D}[g] {\cal D}\phi
 \; {\rm e}^{\textstyle -\bar{\mu} \int {\rm d}^2\xi
\sqrt{g} - S_M(\phi,g)} \;,
\label{eq211}
\end{equation}
where $\bar{\mu}$ is the cosmological constant, the integration is
over equivalence classes of metrics $[g]$, and $S_M(\phi,g)$ is 
the matter Lagrangian.  For a free bosonic string, embedded in
$D$--dimensions, $S_M = \frac{1}{8\pi} \int {\rm d}^2\xi
\sqrt{g} g^{ab} \partial_a \vec{X} \partial_b \vec{X}$, where
$\vec{X}$ are the embedding coordinates of the string.

Discretizing the model Eq.~(\ref{eq211}) using the simplicial
gravity approach \cite{generic}, alias dynamical triangulations, 
the integration
over metrics is replaced by a summation over all triangulations
$T$ constructed by gluing equilateral 
triangles together along their edges
into a simplicial manifold of a given topology \cite{generic}.  
The discretized (grand-canonical) partition function becomes
\begin{equation}
Z(\mu) \;=\; \sum_A {\rm e}^{\textstyle -\mu A}
\; \sum_{T\in \{{\cal T}\}}  \;\int
{\rm d[{\bf x}]} \; \delta({\rm \bf x}_{cm})
\; {\rm e}^{\textstyle
-\sum_{\langle ij\rangle} ({\rm \bf x}_i - {\rm \bf x}_j)^2 },
\label{eq212}
\end{equation}
where $A$ is the area of the triangulation (number of vertices), 
${\rm \bf x}$ is the embedding of a vertex in a $D$--dimensional 
space, and $\langle ij\rangle$ indicates that the sum is 
over adjacent vertices in the triangulation.
The center of mass ${\rm \bf x}_{cm}$ is kept fixed to
eliminate the translational zero mode.

The integration over the Gaussian fields in Eq.~(\ref{eq212})
can be carried out explicitly and the canonical (fixed area) 
partition function becomes
\begin{equation}
 Z_A \;=\; \sum_{T\in{\cal T}(A)} ({\rm det} \; C_T)^{-D/2} \;,
 \label{eq213}
\end{equation}
where $C_T$ is the adjacency matrix of the  the triangulation $T$:
\begin{equation}
 C_T \;=\; \left \{  
 \begin{array}{ll}
  q_i  & \qquad \text{if $i=j$,} \\
  -c_{ij}  & \qquad \text{if $i$ and $j$ are adjacent,} \\
  0     & \qquad \text{otherwise.}
 \end{array}
 \right .
\end{equation}
Here $q_i$ is the order of vertex $i$ and $c_{ij}$ is
the number of edges connecting the adjacent vertices $i$ and
$j$.  Note that as we use degenerate triangulations,
which are defined below, there can
be more than two edges connecting the vertices $i$ and $j$ and
the order of vertices is defined excluding self-loops.
In calculating the determinant of $C_T$ one vertex is 
excluded from the graph in order to eliminate the zero mode.

In the partition function Eq.~(\ref{eq213}) the embedding
dimension $D$ now appears as a free parameter and is no
longer restricted to positive integer values.  In particular,
it can be taken to be negative; this corresponds to coupling
a conformal field theory with a negative central charge $c = D$
to two-dimensional gravity.

The sum over triangulations $T$ in Eq.~(\ref{eq212}) is over an
appropriate class of triangulations ${\cal T}$ of area $A$ and with
fixed topology.
Different classes amount to different discretization of the
manifolds. By universality arguments different choices of ${\cal T}$
should yield the same critical behavior as long as they only differ at
the level of the discretization.  This statement is known to be true
for two commonly used classes of triangulations, {\it degenerate}
(${\cal T}_D$) and {\it combinatorial} (${\cal T}_C$) triangulations.
Combinatorial triangulations are defined by forbidding self-loops (an
edge starting and ending at the same vertex) and vertices connected by
more than one edge.  These pathologies are, on the other
hand, allowed in the class of degenerate triangulations.  For these
particular classes of triangulations the model Eq.~(\ref{eq212}) has
been solved for some special values of $c$ \cite{matis}.
This universality holds even if 
curvature fluctuations of the triangulation, i.e.\ the vertex orders
$q_i$, are maximally restricted, to $q_i =6 \pm 1$,
as shown by numerical simulations \cite{minimal}.

Finite-size effects do, however, depend on the discretization;
in particular, for degenerate triangulations they are substantially
smaller (one order of magnitude) than for combinatorial 
~\cite{mewexl}. 
In the work presented in this paper we use  
degenerate triangulations of spherical topology.

It is possible to get some insight into the behavior of 
the model in the limits $D \rightarrow \pm
\infty$ by looking at what triangulations dominate
the sum in Eq.~(\ref{eq213}) \cite{edgeflip}.  
As the determinant of $C_T$ is 
related to the number of spanning trees on a particular graph,
the graphs dominating in the limit $D \rightarrow +\infty$
are those with a minimal number of spanning trees.
This corresponds to triangulations with a branched polymer
structure. 
In the limit $c \rightarrow - \infty$, on the other hand,
 we expect the determinant $C_T$ to take its largest value for 
triangulations with a maximal number of spanning trees.
This implies that the dominant 
triangulations in this limit will be flat with $q_i = 6$
(apart from few defects $q_i \neq 6$ needed to form a
closed spherical surface). 

\subsection{Simulating a {\boldmath $c=-5$} gravity theory}

We have studied numerically the partition function 
Eq.~(\ref{eq213}) for $c = -5$ using Monte Carlo simulations.  
The space of all triangulations is
explored using the so-called edge-flip algorithm \cite{edgeflip} in
which an edge $l_{ij}$, common to two triangles $t_{ijh}$ and
$t_{kji}$, is removed (or flipped) and replaced by the edge $l_{hk}$.
This algorithm is known to be ergodic (for fixed area).
Unfortunately, the determinant in Eq.~(\ref{eq213}) corresponds to a
non-local action making the simulations very difficult.  Every
time a flip of an edge is proposed the whole determinant has to be
recalculated, an operation requiring on the order of $N^3$ floating point
operations. We can exploit the locality of the flip and using standard
techniques \cite{det} to reduce this to $N^2$ floating point operations
per flip, but this still limited our simulations to triangulations
consisting of up to 1600 vertices.  Nevertheless, those triangulations
are considerably larger than have been simulated before with this
method \cite{detsim}.

\setlength{\tabcolsep}{12pt}
\begin{table}
 \caption{\label{numgr}\small The number $N_g$ of $c=-5$ graphs,
  of area $A$, we have generated by simulating 
  the model Eq.~(\ref{eq213}).  Also shown is the number 
  $N_r$ of different replicas (graphs), for each area,   
  on which we simulated the Ising model (Section~3).}
 \vspace{4pt}
 {\small
 \begin{center}
 \begin{tabular}{|r|r|r|} \hline
  $A$  &  $N_g$ &   $N_r$  \\ 
  \vspace{-10pt}    & &    \\ \hline
  \vspace{-10pt}    & &    \\
  100  &  3000     &  200  \\
  150  &  3000     &       \\
  200  &  3000     &  200  \\
  300  &  3000     &       \\
  400  &  2000     &  130  \\
  600  &  2500     &       \\
  800  &  700      &  60   \\
  1200 &  196      &  178  \\
  1600 &  96       &  96   \\\hline  
 \end{tabular}
 \end{center} }
\end{table}
 
In the simulations we stored triangulations after every 10 to
20 Monte Carlo sweeps, were each sweep consisted of
flipping, approximately, an area worth of edges.  With the
auto-correlations present, which are relatively modest on
such small surfaces, the stored graphs were more or less
independent.  In Table~1 we show the total number of 
graphs we generated, and analyzed, for each area.

\subsection{The string susceptibility exponent 
            {\boldmath $\gamma_s$}}

To determine the critical behavior of the model Eq.~(\ref{eq213}),
with $c = -5$, we measured two critical exponents:
the {\it string susceptibility} exponent $\gamma_s$ 
and the {\it fractal} or Hausdorff dimension $d_H$.
The latter will be discussed in the next section, here we
consider $\gamma_s$.

The string susceptibility exponent is defined by the singular
behavior of the grand-canonical partition function Eq.~(\ref{eq212})
as the cosmological constant $\mu$ approaches its critical value:
\begin{equation}
 Z(\mu) \;\approx\; Z_{\rm reg} \;+\; (\mu - \mu_c)^{2 - \gamma_s}.
 \label{freesing}
\end{equation}
This implies that the canonical partition function behaves 
asymptotically as
\begin{eqnarray}
 Z(A) &\sim& {\rm e}^{\textstyle \; \mu_c A} A^{\gamma_s - 3} \;,
 \qquad A \rightarrow \infty \;.
 \label{asymgam}
\end{eqnarray}
The value of $\gamma_s$ can be calculated from the Liouville field
theory \cite{kpz}:
\begin{equation}
 \gamma_s \;=\; \frac{1}{12} \left (c - 1 -
 \sqrt{(c-25)(c-1)} \right).
 \label{kpz}
\end{equation}
For $c>1$ this implies a complex critical exponent and the
corresponding theory is not well defined ---  this is related 
to the existence of tachyons in string theories in 
embedding dimension larger than two.
For $c \leq 1$, on the other hand, the theory is well defined and,
specifically, for $c=-5$ Eq.~(\ref{kpz}) predicts 
$\gamma_s = -1.618\ldots$  

In order to measure $\gamma_s$ for 
the ensemble of triangulations, generated in our
simulations, we study the distribution of so-called
{\it minbus} (minimal neck baby universes) \cite{jain}.  
A minbu is a part of the triangulation
connected to the rest through a minimal neck; for a degenerate
triangulation a minimal neck is simply a self-loop.
By counting in how many ways a minbu of size $B$ can
be connected to a surface of size ($A-B$), the
size distribution of minbus can be written as:
\begin{eqnarray}
 n_A(B) &\;=\; &\frac{\textstyle B \; Z(B) \; (A-B) 
 \; Z(A-B)}{\textstyle Z(A)}  \\
 &\;\sim\; &\left [ (A-B) \; B\right ]^{\gamma_s - 2}.
 \label{minbu}
\end{eqnarray}
In the last step we have used the asymptotic behavior
of the partition function, Eq.~(\ref{asymgam}) --- in this way
the leading exponential behavior cancels out and we
can directly access the sub-leading corrections governed
by $\gamma_s$.

The distribution $n_A(B)$ is easily measured in numerical
simulations and $\gamma_s$ extracted by a fit to Eq.~(\ref{minbu}).
In practice, though,  one has to impose a lower cut-off on
the size of minbus included in the fit as there are finite-size 
corrections to the asymptotic form Eq.~(\ref{asymgam}).  
This can be done in a systematic way;
small minbus are thrown away in the fitting procedure
until one gets a stable value of $\gamma_s$ and an acceptable
$\chi^2$--value for the fit \cite{gammeas}.  
Unfortunately, the large negative 
value of $\gamma_s$ for the $c=-5$ graphs
makes this measurements more difficult than for previously
studied ensembles of triangulations,
as the distribution falls off very rapidly. 

We have measured $\gamma_s$ for surfaces of area up to 300 vertices;
for the larger surfaces our statistics was not sufficient for a
reliable determination of the minbu distribution.  Although these are
very small surfaces, our experience in measuring $\gamma_s$ for other
models of dynamical triangulations shows that they are large enough,
provided one is using degenerate triangulations \cite{mewexl}.  In this
particular case the lack of statistics is much more of a problem than
the smallness of the triangulations.  
But as determining $\gamma_s$ is of
secondary interest for us, we did not deem it worthwhile 
to invest too much CPU--power in increasing the statistics.  
We get the following values of $\gamma_s$:
\begin{center}
\begin{tabular}{|c|l|}\hline
 $A$ &  $\gamma_s$  \\ \hline
 100 &  -1.52(17)   \\
 150 &  -1.42(15)   \\
 200 &  -1.37(22)   \\
 300 &  -1.71(19)   \\ \hline
\end{tabular}
\end{center}
Those values are
in reasonable agreement with the theoretical value 
$\gamma_s = -1.618\ldots$
 
\subsection{The vertex-vertex correlation function and 
            {\boldmath $d_H$}}

Another exponent that characterizes the fractal structure of
the surfaces is the (intrinsic) fractal dimension $d_H$.
It can be defined from the behavior of the vertex-vertex
(or two-point) correlation function $g_A(r)$:
\begin{equation}
 g_A(r) \;=\; \frac{1}{A} \; \langle  \sum_{i,j} 
  \delta(d_{ij} - r) \rangle_T \;, 
 \label{vercor}
\end{equation}
where $d_{ij}$ is the (minimal) geodesic distance between two
vertices $i$ and $j$ and the statistical average is performed 
over all triangulations $T$.  This correlation function
simply counts the number of vertices (or the area of the manifold) 
at a geodesic distance $r$ from a marked vertex $i$, averaged over
all vertices $i$.  We expect its short distance 
behavior to be $\;g_A(r) \sim r^{d_H - 1}$, provided $r \ll A^{1/d_H}$.

On a triangulation we define the geodesic distance between
two vertices as the shortest path between them traversed
along links.  Alternatively, one can define a 
triangle-triangle correlation function $t_A(r)$, analogous
to Eq.~(\ref{vercor}), in which case the geodesic distance
is defined as the shortest path between two triangles
traversed through the center of triangles.
Although those two definitions will result in very different
correlation functions for a particular triangulation, they 
should define the same fractal dimension. 
  
To extract the fractal dimension from the measurements of
the correlation function Eq.~(\ref{vercor}) it is convenient
to use methods of finite-size scaling. 
Assuming that the only relevant length-scale in the model is 
defined by $A^{1/d_H}$, general scaling arguments 
\cite{hausd,janhaus} imply that
\begin{equation}
 g_A(r) \;\sim\; A^{1-1/d_H} \; F(x),
 \label{dhscal}
\end{equation}
where we have introduced the scaling variable 
\begin{equation}
 x = \frac{r}{A^{1/d_H}} \,.
 \label{scalvar}
\end{equation}
In fixing the prefactor in Eq.~(\ref{dhscal}) 
we have used that $\sum_r g_A(r) = A$.

In practice there are strong finite-size corrections to 
this scaling behavior.  The measurements of $d_H$ can be 
improved, considerably, by including the simplest finite-size 
correction to $x$, i.e.\ by introducing the so-called
{\it shift} \cite{janhaus}:
\begin{equation} 
 x = \frac{r+a}{A^{1/d_H}} \,
 \label{shift}
\end{equation}
where the shift parameter $a$ will in general depend on 
the particular observable we consider. 
This scaling correction has been applied successfully to
all geometric and matter correlation functions, even with 
quite different functional dependence on $x$ \cite{amko}. 
Its possible geometric
origin has been investigated analytically for $c=0$ in \cite{cm2}.
The shift was also found and calculated analytically in
correlation functions on branched polymers where it has been shown to
contain the singularity responsible for the crumpling phase
transition \cite{bb}.

It is possible to generalize Eq.~(\ref{vercor}) to higher
moments of the correlation function.  Introducing the 
loop-length distribution $\rho_A(r,l)$, which 
counts the number of loops of length $l$ at geodesic distance
$r$ from a marked vertex $i$, we define the $k$--th 
moment of $l$ as \cite{anhaus2,amko,cm2}
\begin{equation}
 l^k_A(r) \;=\; \sum_l l^k \; \rho_A(r,l) \;.
 \label{moment}
\end{equation}
Note that $g_A(r) = l^1_A(r)$.  On a triangulation loops
are defined as a connected (by links) subset of the vertices 
that are at distance $r$ from $i$.\footnote{In a similar way
we can define a loop-length distribution for the
triangles at a distance $r$ from a marked triangle.  In that case
we define a loop as a connected subset of triangles sharing
at least one vertex.  If we define loops as triangles
sharing at least one link, then those loops always stay at the
level of the lattice cut-off.}  However, the generalization of
the scaling hypothesis Eq.~(\ref{dhscal}) is more subtle,
as it depends on how the boundary length $l$ scales with
the radius $r$ \cite{cm2}.  If we assume that 
${\rm dim}[r^2] \equiv {\rm dim}[l] = {\rm dim}[A^{2/d_H}]$, 
then we have for the moments:
\begin{equation}
 l^k_A(r) \;=\; A^{2k/d_H} \; F_k(x) \,, \qquad k > 1.
 \label{momentscal}
\end{equation}
Alternatively, if we assume that 
${\rm dim}[l^2] \equiv {\rm dim}[A] = {\rm dim}[r^{d_H}]$, we get:
\begin{equation}
 l^k_A(r) \;=\; A^{k/2} \; \tilde{F}_k(x).
 \label{momentscal2}
\end{equation}
Our measurements clearly favor the former scaling form 
Eq.~(\ref{momentscal}); this is compatible with results recently
obtained for a $c=-2$ theory \cite{cm2}.
This implies, from dimensional analysis, 
that the boundary length $l$ acquires an 
anomalous scaling dimension with the area;
at present the origin of this behavior is still not understood.  
Note that in the only soluble case, $c=0$, 
both scaling forms agree. 

We have measured the moments $l^k_A(r)$ for $k \leq 4$,
together with the triangle-triangle correlation function
$t_A(r)$.  The scaling behavior, 
Eqs.~(\ref{dhscal}) and (\ref{momentscal}),
is analyzed by ``collapsing'' the distributions
corresponding to different area onto a single curve.  For
this purpose it is convenient to interpolate between the
(discrete) values of $x$ --- this we do using a
cubic-spline (see Ref.~\cite{cm2} for details).  
The collapse depends on only two free
parameters, $a$ and $d_H$; their optimal values are determined
by minimizing the $\chi^2$--value for the collapse, where the 
$\chi^2$--value is
defined by the difference between the curves after rescaling.
The results are shown in Table~2. We used 
triangulations of area 200 to 1600 in the analysis;
if we included smaller graphs it was not possible to
get an acceptable $\chi^2$--value for the fit.  
We observe, as has been observed before \cite{hausd,janhaus}, 
that the finite-size effects are considerable
bigger for the triangle-triangle correlation function.
This is also reflected in much larger value of the 
shift parameter $a$.

\setlength{\tabcolsep}{12pt}
\begin{table}
 \label{table2}
 \caption{\small The fractal dimension $d_H$ for an ensemble of
  $c=-5$ graphs, determined by collapsing the correlation functions
  $l^k_A(r)$ onto a single curve using the
  scaling forms Eqs.~(\ref{dhscal}) and (\ref{momentscal})
   and the shift Eq.~(\ref{shift}).
  The corresponding (optimal) value of the shift parameter $a$
  is included.  Graphs of area 200 to 1600 were included in the
  analysis.  The corresponding values for the triangle-triangle
  correlation function $t_A(r)$ are also included.}
  \vspace{8pt} 
  \begin{center}
  \begin{tabular}{|cc|ll|} \hline
   \rule[-2.5mm]{0mm}{7mm} $l_A^k(r)$ &   $k$   &   $d_H$   &   $a$   \\\hline
   \rule{0mm}{5mm} $g_A(r)$ &  1     &   3.36(16)   &  0.6(6)   \\
   &              2     &   3.33(5)    &  0.4(3)   \\
   &              3     &   3.36(4)    &  0.4(3)   \\
  \rule[-2.5mm]{0pt}{0mm} &              4     &   3.36(4)    &  0.4(3)   \\
  \rule[-2.5mm]{0pt}{0mm}    $t_A(r)$       &   
    &   3.07(24)   &  2.4(1.6) \\ \hline
  \end{tabular}
  \end{center}
\end{table}

\begin{figure}
\epsfxsize=4in \centerline{ \epsfbox{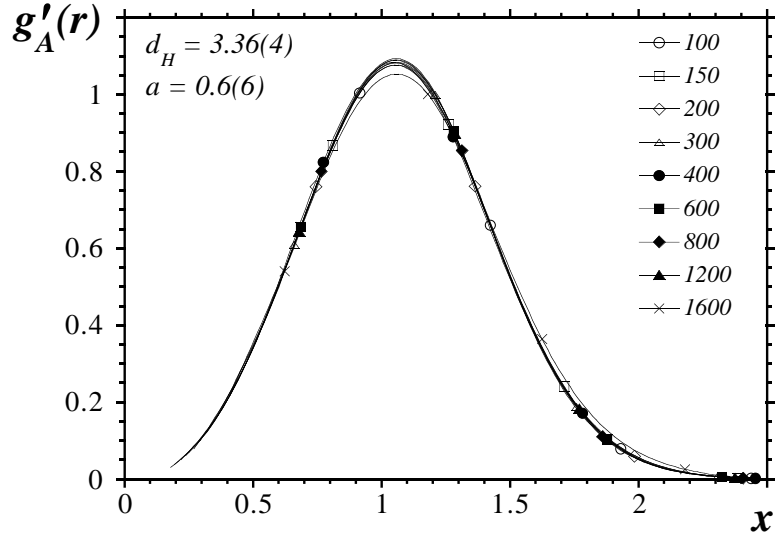}}
\caption{\small The scaled vertex-vertex correlation function
 $g^{\prime}_A(r) \equiv g_A(r)/A^{1-1/d_H}$ {\it vs}.\ the
 scaling variable $x = (r+a)/A^{1/d_H}$.  The measured value
 of the fractal dimension $d_H = 3.36$ and a shift $a=0.6$
 were used in the scaling.}
\label{fighausscal}
\end{figure}

To demonstrate how good this scaling behavior actually is, we show
in Figure~1 the scaled curves for the
vertex-vertex correlation function $g_A(r)$.  These curves have been
scaled using $d_H = 3.36(4)$, the average value of the fractal
dimension from Table~2, and the corresponding
value of the shift.  The quality of the scaling for the
higher moments, and for $t_A(r)$, is equally impressive.

Finally, we look at the loop-length distribution $\rho_A(r,l)$.
Although, in principle, it contains the complete
set of informations about the moments, it is not very
convenient for extracting the fractal dimension.
Nevertheless, we demonstrate that its scaling behavior
is consistent with the measured fractal dimension.
Since our measurements show that for the higher moments
$l \sim r^2$ (see Eq.~(\ref{momentscal})), i.e.\ the
same behavior as for $c=0$, we expect that
\begin{equation}
\label{rhosc}
\rho_A(r,l)  \;\approx \; r^2 F_{\rho}(l/r^2).
\end{equation}
We show an example of this scaling in Figure~2. The relation
Eq.~(\ref{rhosc}) is, of course,
exact only for $\rho_\infty(r,l)$, as is the case for $c=0$
\cite{anhaus2}, and finite-size effects are expected for small
$l/r^2$. The deviation from the above scaling behavior is,
in principle, determined by the scaling corrections 
Eqs.~(\ref{dhscal}) and (\ref{momentscal}) \cite{amko}.

\begin{figure}
\epsfxsize=4in \centerline{ \epsfbox{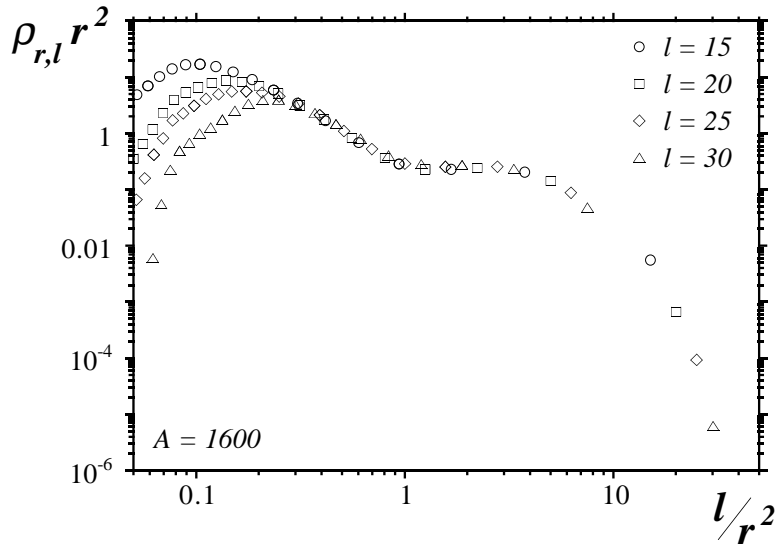}}
\caption{\small 
  An example of the scaling behavior of the loop-length
  distribution function $\rho_A(r,l)$ for $A=1600$.
  The distributions for different choices of $l$ are re-scaled
  in accordance with Eq.~(\protect\ref{rhosc}).}
\label{figloopscal}
\end{figure}

As we mentioned in the Introduction, there exist several theoretical 
predictions for the fractal dimension for theories 
of two-dimensional quantum gravity coupled to matter.
Using string field theory,
or a transfer matrix approach, with a modified definition
of a geodesic distance, one gets \cite{anhaus1}:
\begin{equation}
 \label{andH1}
 d_H \;=\; \frac{2}{|\gamma_s|}  \;=\;
  \frac{24}{1 - c + \sqrt{(25-c)(1-c)}} \;.
\end{equation}
An alternative expression is obtained by studying the diffusion on the
surface, within the framework of Liouville theory \cite{anhaus2}:
\begin{equation}
 \label{andH2}
 d_H \;=\; 2 \times 
  \frac{\sqrt{25 -c} \;+\; \sqrt{49-c}}
       {\sqrt{25-c}  \;+\: \sqrt{1-c}} \;.
\end{equation}
These two analytical predictions disagree, except in the
the case of pure gravity; the only case where an
exact solution exist \cite{anapg}.
For $c=0$ we have $d_H=4$, this value is also obtained in
numerical simulations using the scaling methods described 
above \cite{hausd,janhaus}.  On the other hand, both Eqs.~(\ref{andH1})
and (\ref{andH2}) disagree with the results of extensive 
numerical simulations of conformal field theories with
$0 < c \leq 1$ coupled to gravity \cite{hausd,janhaus,aamt,amko}.  
The numerical simulations 
indicate that the fractal dimension is 4 independent of 
the coupling to matter.
In \cite{hausmjg}, the discrepancy for $c = 1/2$ --- one Ising model
coupled to gravity --- has been attributed to the subtlety in how the
continuum limit should be taken is this model. An alternative proposal
has been made in \cite{aajk} where the authors suggest that a
different scaling variable, related to the size of clusters, should be
used in the derivation of Eq.~(\ref{andH1}) instead of the geodesic
distance. It should be
emphasized, however, that the present numerical accuracy is not enough
to rule out completely the prediction of Eq.~(\ref{andH2}) for $c >
0$.

The only model with $c<0$ where the fractal dimension has
been measured numerically is $c=-2$ or topological gravity
\cite{cm2}.  This is a very special case were it is
possible to sample the space of triangulations
recursively; this allow simulations of very large graphs 
(graphs of up to $8\times10^6$ triangles were
used in Ref.~\cite{cm2}).  The measured fractal
dimension in this case is $d_H = 3.58(4)$, which agrees
very well with the prediction from Liouville theory, 
Eq.~(\ref{andH2}): $d_H = 3.561\ldots$

Our result for $c=-5$, $d_H = 3.36(4)$, also agrees 
reasonably well with Eq.~(\ref{andH2}), which predicts
$d_H =  3.236\ldots$ for $c=-5$, especially given the smallness
of our graphs.  On the other hand, it completely
rules out the prediction of Eq.~(\ref{andH1}): $d_H \approx 1.236$.
Although the value 3.36(4) is slightly larger than
predicted by Eq.~(\ref{andH2}), we notice that the
fractal dimension obtained using the triangle-triangle 
correlation function: $d_H = 3.07(24)$, is somewhat smaller .
This effect was also noticed for the $c=-2$ model where
the two values converged to Eq.~(\ref{andH1})
on larger graphs.  This is what we observe in our simulations
as well.

\subsection{Comparison with other random lattices}

\noindent
The fractal structure of the $c=-5$ graphs
differ substantially from other types of random 
lattices frequently studied.
An ensemble of random lattices, commonly used in the
study of quenched disorder, is Poissonian random lattices.
They are constructed by distributing vertices 
uniformly on a two-dimensional
manifold and link them together to form a triangulation, 
usually following a prescription by Dirichlet and Voronoi 
\cite{itzy}.

We have compared the properties of our ensemble of 
graphs to that of Voronoi triangulations by looking at the 
probability distribution of vertex orders $p_n$.
In Figure~3 we plot this distribution for the 
$c = -5$ graphs, Voronoi triangulations, and, for comparison,
pure gravity ($c=0$) and branched polymer ($c=5$) graphs.  
As in all cases the distributions are for triangulations, 
they all have the mean value 
$\bar{p}_n = 6$, but in other aspects they differ.
The distribution for Voronoi graphs is peaked sharply around the
mean value and falls off rapidly as $n$ increases.
The pure gravity distribution peaks at the smallest
possible curvature ($n=1$ for degenerate triangulations) 
and falls off much slower.
The distribution for $c=-5$ lies in-between, as could
be expected from its fractal dimension\footnote{It is worth 
noting that the fractal dimension alone is not enough
to characterize the graphs. For example, both a flat 
lattice and branched polymers have $d_H = 2$}.

\begin{figure}
\epsfxsize=4in \centerline{ \epsfbox{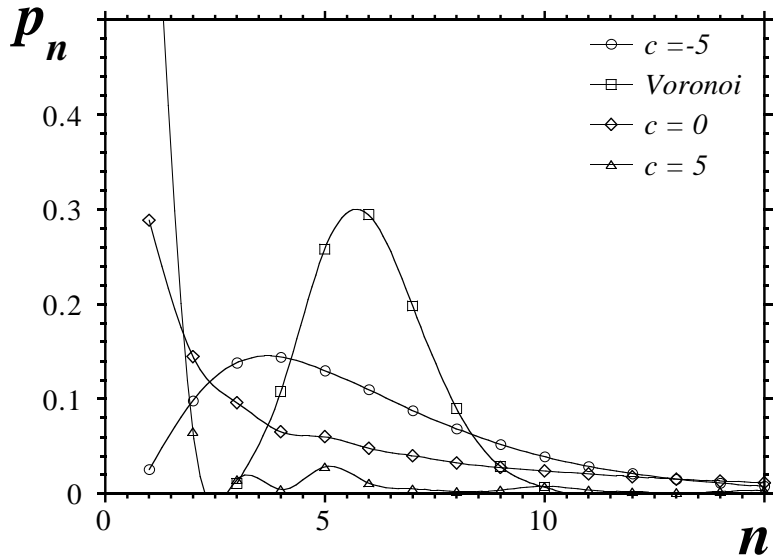}}
\caption{\small The (normalized) curvature distribution
 $p_n$ for an ensemble of graphs corresponding to 
 $2d$--gravity coupled to matter with $c= -5$, 0 and 5, respectively,
 and for Voronoi triangulations.  The interpolating curves
 are just to guide the eye.} 
\label{p_n}
\end{figure}

As Figure~3 indicates, the randomness of the Voronoi
triangulations is only at the local level, their global
fractal structure, defined by, for example, the fractal 
dimension or the string susceptibility exponent, will
be the same as of the underlying two-dimensional flat lattice
used in constructing them.  In this respect their randomness
is only a (small) local perturbation around the flat background.
This is also reflected in that the critical behavior of spin
models on such triangulations is the same as for a flat
lattice \cite{jankevor}.
This is contrary to the other ensembles of graphs
shown in Figure~3, which have a genuinely different
global fractal structure.

It should also be emphasized that although local observables 
such as $p_n$ reveal something about the randomness
of the graphs, they are not a universal 
property and not a good indicator for the critical behavior
of the corresponding model.
Within the context of $2d$ gravity the local properties
of a given model can be changed by adding an
irrelevant operator to the action in Eq.~(\ref{eq212}); for example,
a term that couples to the local curvature.  
But, except in extreme cases, this modification
does not affect the critical behavior of the model
\cite{minimal,kaz}. 

It is, however, possible to change the global fractal properties
of the graphs defined by the $2d$-gravity model Eq.~(\ref{eq213})
by adjusting the embedding dimension $D$.  In particular,
by taking $D$ negative we can create an ensemble of
triangulations with any fractal dimension between 2 and 4.
This makes the model of Eq.~(\ref{eq213}) ideal for investigating how the
fractal properties, like the fractal dimension, affect
the critical behavior of a spin model living on the 
graphs.  As will be demonstrated in the next section, 
in the case of the Ising model this results in critical
exponents that are radically different from those of the
Ising model on a flat lattice.

\section{The Ising model on {\boldmath $c=-5$} graphs}

\subsection{Predictions from the continuum}

In this second part of the paper we investigate how the critical
behavior of an Ising model is modified when it is defined on a
quenched ensemble of $c=-5$ graphs.  In the annealed case coupling a
conformal field theory to $2d$-gravity results in dressing the
dimensions of primary operators of the theory. The dressed weights are
given by the KPZ formula \cite{kpz}.
From those weights one can calculate the new critical exponents.
As a dynamical triangulation, the only case that has
been solved explicitly is the one Ising model coupled
to gravity, formulated as a two-matrix model \cite{matis}.  
In that case the phase transition changes from second to
third order and the critical exponents agree with
the ones calculated using the KPZ formula. 

All these calculations are, as emphasized above, for the Ising model
on an annealed ensemble of graphs.  Now the conjecture, discussed in
the Introduction, is that the interaction between the Ising model and the
geometry is not important; the critical behavior is simply determined
by the average fractal structure of the triangulations.  This implies
that we can also use the KPZ formula to calculate the critical
behavior for an Ising model on a quenched ensemble of graphs,
provided they define the appropriate fractal structure.  If the
conjecture is true the critical behavior of an Ising model on a
quenched ensemble of $c=-5$ graphs is the same as that of the Ising
model 
coupled dynamically to gravity with additional $c=-11/2$
conformal field coupled to it. 

Given a primary operator
of conformal weight $\Delta^0$ in the original theory,
the KPZ formula gives its conformal weight after 
coupling to gravity:
\begin{equation}
\Delta \;=\; 2 \times \frac{\sqrt{1-c+12\Delta^0}
  -\sqrt{1-c}}{\sqrt{25-c}-\sqrt{1-c}} \;.
 \label{dressing}
\end{equation}
For the Ising model the relevant operators are the
energy density $\epsilon$ and the spin $\sigma$; 
from the conformal weight of those operators
we can calculate the specific heat exponent $\alpha$ and the 
magnetization exponent $\beta$:
\begin{equation}
\alpha \;=\; \frac{2(1-\Delta_{\epsilon})}{2-\Delta_{\epsilon}}
\qquad {\rm and} \qquad
\beta \;=\; \frac{\Delta_{\sigma}}{2-\Delta_{\epsilon}}.
\label{isexp}
\end{equation}
In flat space the conformal weights of the Ising operators 
are: $\Delta_{\epsilon}^0 = 1$ and
$\Delta_{\sigma}^0 = \frac{1}{8}$; hence
the KPZ formula, together with Eq.~(\ref{isexp}),
predicts $\alpha \approx -0.452432$ and $\beta \approx 0.234186$
for an Ising model on $c=-5$ graphs.
The full list of exponents is shown in Table~5.
As those exponents are very different from both the
Ising model exponents on a flat lattice (the Onsager exponents) 
and coupled dynamically to pure gravity (both sets of exponents 
are included in the table)
it should be relatively easy to distinguish them in numerical
simulations. 

\afterpage{\clearpage}

\subsection{The Ising model simulations}

We have simulated the Ising model for several independent replicas
of the $c=-5$ graphs we generated.  We did this for graphs of size
100, 200, 400, 800, 1200 and 1600 vertices --- 
for the number of replicas we sampled see Table~1.
We use the standard definition of the Ising model:
\begin{equation}
Z_{\rm Ising}(T) \;=\; \sum_{\{\sigma_i\}}
 {\rm e}^{\textstyle \; \beta \sum_{\left < ij \right >}
 \sigma_i \sigma_j }\;,
\label{ising}
\end{equation}
where $\sigma_i=\pm1$ and $\left<ij\right>$ denotes 
adjacent vertices in the graph $T$.  The Ising spins are placed
on the vertices of the graphs; alternatively they could be put
on the triangles, but based on duality arguments 
we would expect the same critical behavior
(this is the case for an Ising model coupled dynamically
to $2d$--gravity \cite{matis}).  
Naively, one might expect that placing the
spins on the triangles was preferable as for a graph of a given
area this yield twice as many spins. But as the finite-size effects
are dominated by the geometry not the number of spins 
this not the case.  In fact, it would only
increase the computational efforts in updating the spin
configuration. 

We simulated at several values of the 
inverse temperature $\beta = J/k_BT$ for $\beta \in [0.15,0.25]$.
A Swendsen-Wang cluster algorithm was used to update the spin
configuration and typically about 200.000 measurements taken
at each $\beta$ value; the measurements were separated 
by 10 to 20 cluster updates.
For each measurement we stored the total energy of the system,
$E = \sum_{\left < ij \right >} \sigma_i \sigma_j$, and the
magnetization, $M$~=~$\sum_i \sigma_i$; all other observables
can be constructed from the those two.  To interpolate between      
measurements at different temperatures we used standard
multi-histogram methods \cite{histogram}.  

All observables ${\cal O}$ were calculated for 
each replica independently and the average over the different
replicas $r$ performed afterwards:
\begin{equation}
\bar{\cal O} \;=\; \frac{1}{N_r} \sum_r {\cal O}_r \;.
\label{repav}
\end{equation}
This is in accordance with
the philosophy that the quenched average should be performed at
the level of the free energy, {\it not} at the level of the
partition function \cite{bind1}.  The error on $\bar{\cal O}$
is estimated by a jackknife analysis over different
replicas.  

To analyze the critical behavior of the Ising model we 
constructed several standard observables. In addition to
the energy density $\left < e \right >$ and magnetization
$\left < m \right >$ per vertex we considered:
\begin{equation}
\label{eqobs1} 
\begin{array}{llll}
C_A  & =\;  \beta^2 A \left( \langle e^2\rangle 
             - \langle e\rangle^2 \right) 
     & \approx\;  c_0 + c_1 A^{\alpha / \nu d_H}
     &      \text{\small (specific heat)}, 
     \vspace{5pt} \\
\chi & =\;  A \left( \langle m^2 \rangle 
             - \langle m \rangle >^2 \right)
     & \sim\;  A^{\gamma /\nu d_H}  
     &   \text{\small (magnetic suscept.)}, 
     \vspace{5pt}\\ 
U_A  & =\; \frac{\textstyle \langle m\rangle^4}
                {\textstyle \langle m^2\rangle^2}
           & \approx c^{\prime}_0 
           & \text{\small (Binder's cumulant)}, \vspace{5pt}\\
V_A  & =\; \frac{\textstyle \langle e\rangle^4}
                {\textstyle \langle e^2\rangle^2}
           & 
           & \text{\small (energy cumulant)},
\end{array}
\end{equation}
together with the various derivatives of the magnetization:
\begin{equation}
 \begin{array}{llll}
  {\rm D}_{|m|}  & \equiv
  \frac{\textstyle {\rm d}\langle |m| \rangle}
     {\textstyle {\rm d}\beta} & =\;
           A \left( \langle e \rangle \langle m\rangle 
             - \langle e|m|\rangle \right)  
           & \sim A^{(1-\beta)/\nu d_H} \vspace{5pt} \\
  {\rm D}_{\ln |m|}  &\equiv
  \frac{\textstyle {\rm d}\ln \langle |m| \rangle}
     {\textstyle {\rm d}\beta} & =\;
           A \left( \langle e \rangle
             - \frac{\textstyle \langle e|m|\rangle}
                    {\textstyle \langle |m|\rangle} \right)  
           & \sim A^{1/\nu d_H} \vspace{5pt} \\
  {\rm D}_{\ln m^2}  &\equiv
  \frac{\textstyle {\rm d}\ln \langle m^2 \rangle}
     {\textstyle {\rm d}\beta} & =\;
           A \left( \langle e\rangle
             - \frac{\textstyle \langle em^2\rangle}
                    {\textstyle \langle m^2\rangle} \right)  
           & \sim A^{1/\nu d_H} \vspace{5pt} \\
  {\rm D}_{U_A}  & \equiv
  \frac{\textstyle {\rm d} U_A}
     {\textstyle {\rm d}\beta} & =\;
           A(1-U_A) \left ( {\scriptstyle 
            \langle e \rangle - 2 \frac{\langle 
            m^2e \rangle}{\langle m^2\rangle} +
            \frac{\langle m^2e \rangle}{\langle m^4\rangle }
            } \right )  
           & \sim A^{1/\nu d_H}.
 \end{array}
 \label{der}
\end{equation}
In Eqs.~(\ref{eqobs1}) and (\ref{der}) we have 
included the expected finite-size 
scaling behavior. For some of the observables this scaling behavior 
applies both at the infinite-area critical temperature $\bar{\beta}_c$ 
and to the scaling of their extremal values. 


\subsection{The critical temperature {\boldmath $\bar{\beta}_c$}}

\noindent
To determine the critical temperature $\bar{\beta}_c$ we use that
the pseudo-critical temperature $\bar{\beta}_c(A)$, defined
by the location of peaks in the different observables, 
approaches the infinite-area value like:
\begin{equation}
\label{coupscal}
 \bar{\beta}_c(A) \;\approx\; \bar{\beta}_c 
+ \frac{c_0}{A^{1/\nu d_H}} \;.
\end{equation}

\begin{figure}[t]
\epsfxsize=4in \centerline{ \epsfbox{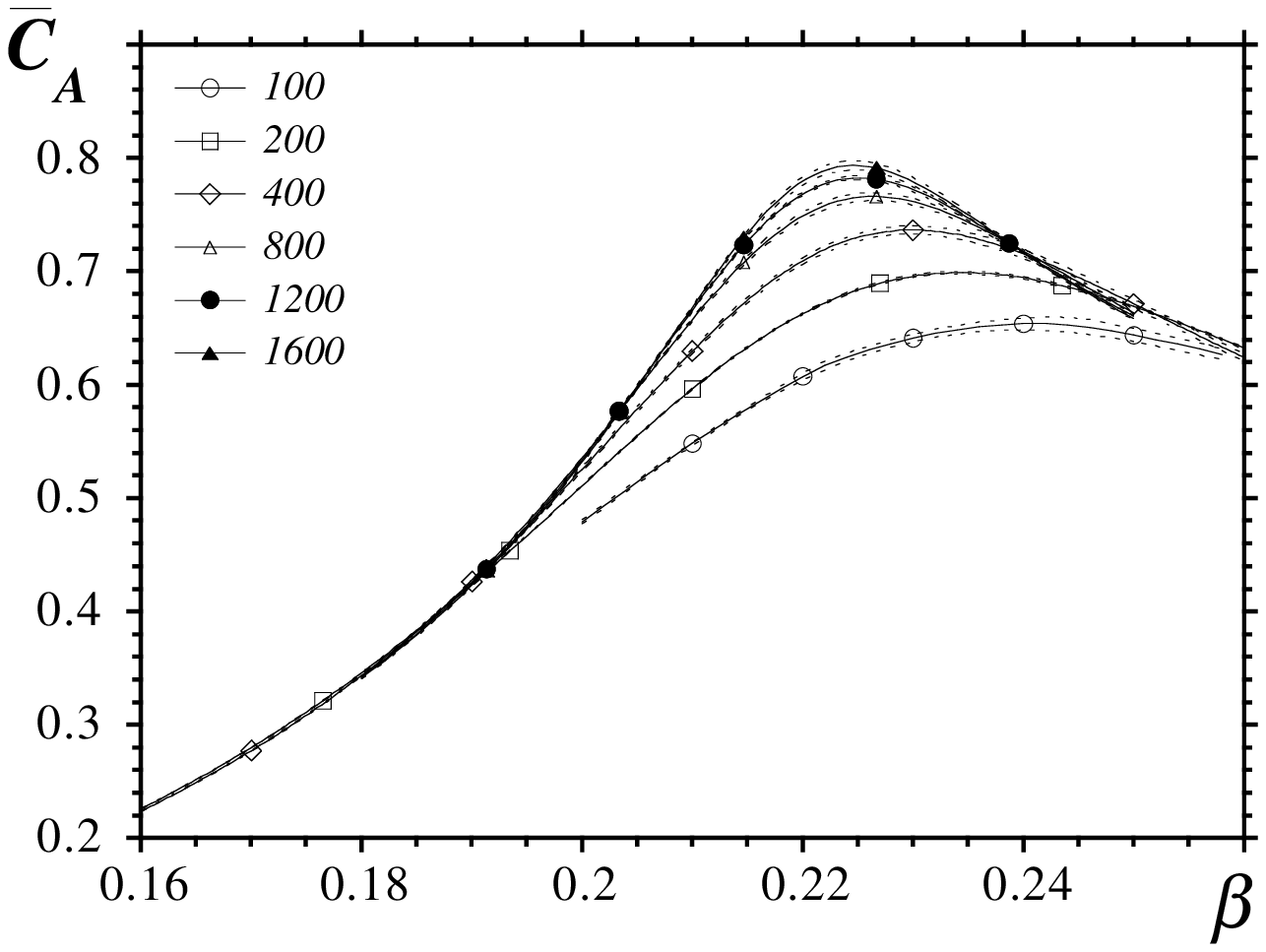}}
\epsfxsize=4in \centerline{ \epsfbox{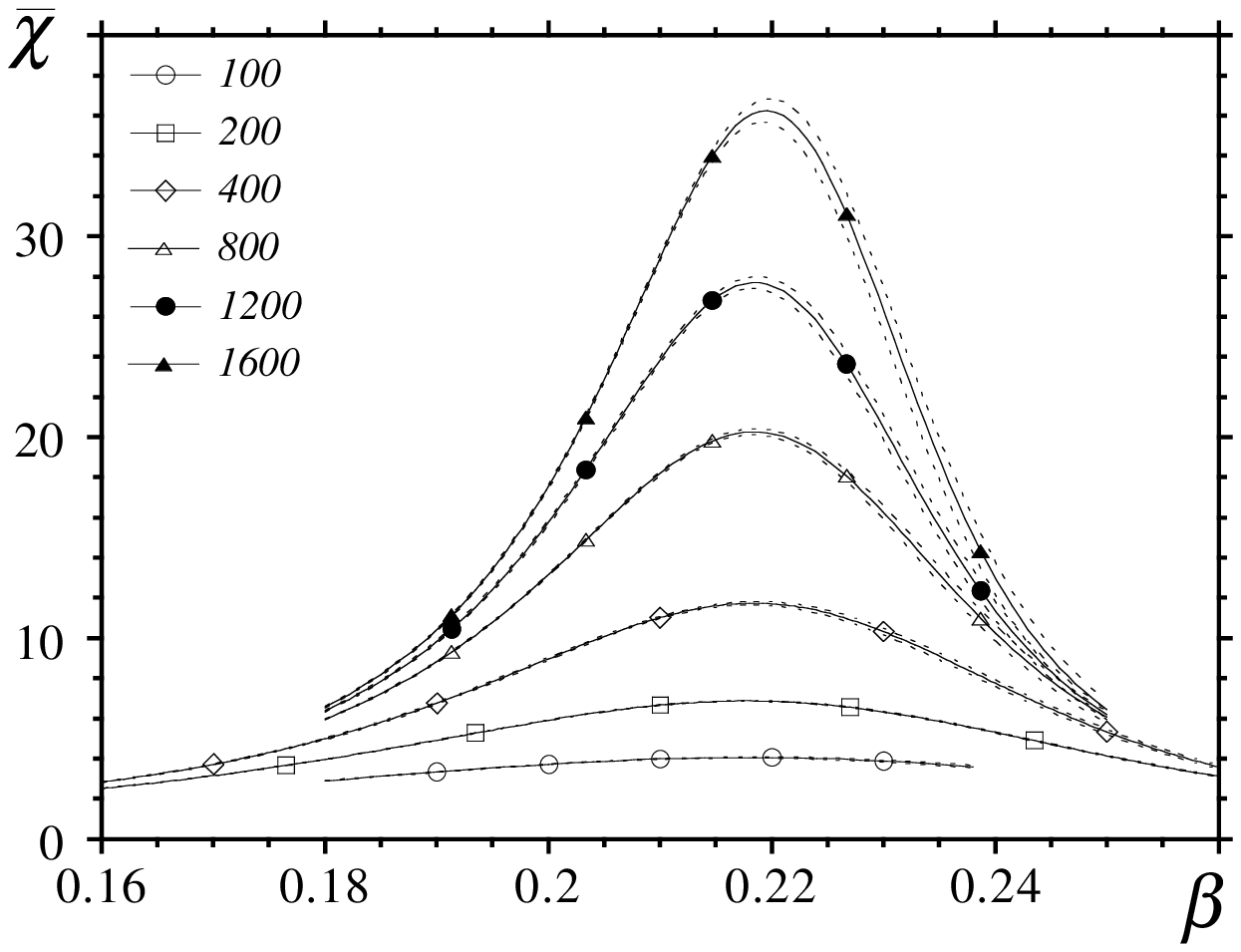}}
\caption{\small ({\it a}) The specific heat $\bar{C}_A$ for an
  Ising model on an ensemble of quenched $c=-5$ graphs.
  ({\it b}) The corresponding magnetic susceptibility $\bar{\chi}$.
  For both observables the curves, for each area $A$, are the
  averages over different replicas and the errors (dashed lines)
  are estimated using jack-knifing over replicas.}
 \label{cvsucc}
\end{figure}

\afterpage{\clearpage}

The observables we have used for this purpose are:
$\bar{C}_A$, $\bar{\chi}$, ${\rm \bar{D}}_{|m|}$, 
$\bar{{\rm D}}_{\ln |m|}$, $\bar{{\rm D}}_{\ln m^2}$,
and $\bar{{\rm D}}_{U_A}$,
all of which have have well resolved peaks.  An example of
this is shown in Figure~4 were we
plot the specific heat and the magnetic
susceptibility.
In the figure the curves shown are the averages over
different replicas of the model, Eq.~(\ref{repav}).
To determine $\bar{\beta}_c(A)$, 
and to estimate the
corresponding error, we locate the peaks on each replica 
independently and then take the averages over the replicas.
In the infinite-area limit, were one expects the triangulations
to become self-averaging with respect to the fractal 
structure, the distribution of $\beta_c^r(A)$ should approach
a delta-function centered at $\bar{\beta}_c$.  This we
demonstrate in Figure~5 where we show the
(normalized) distributions of pseudo-critical
temperatures $\rho(\beta_c^r)$; in this case for the specific heat.  
And, indeed, the distributions get narrower
as the area is increased.  Similar behavior is observed for
the other observables.

\begin{figure}
\epsfxsize=4in \centerline{ \epsfbox{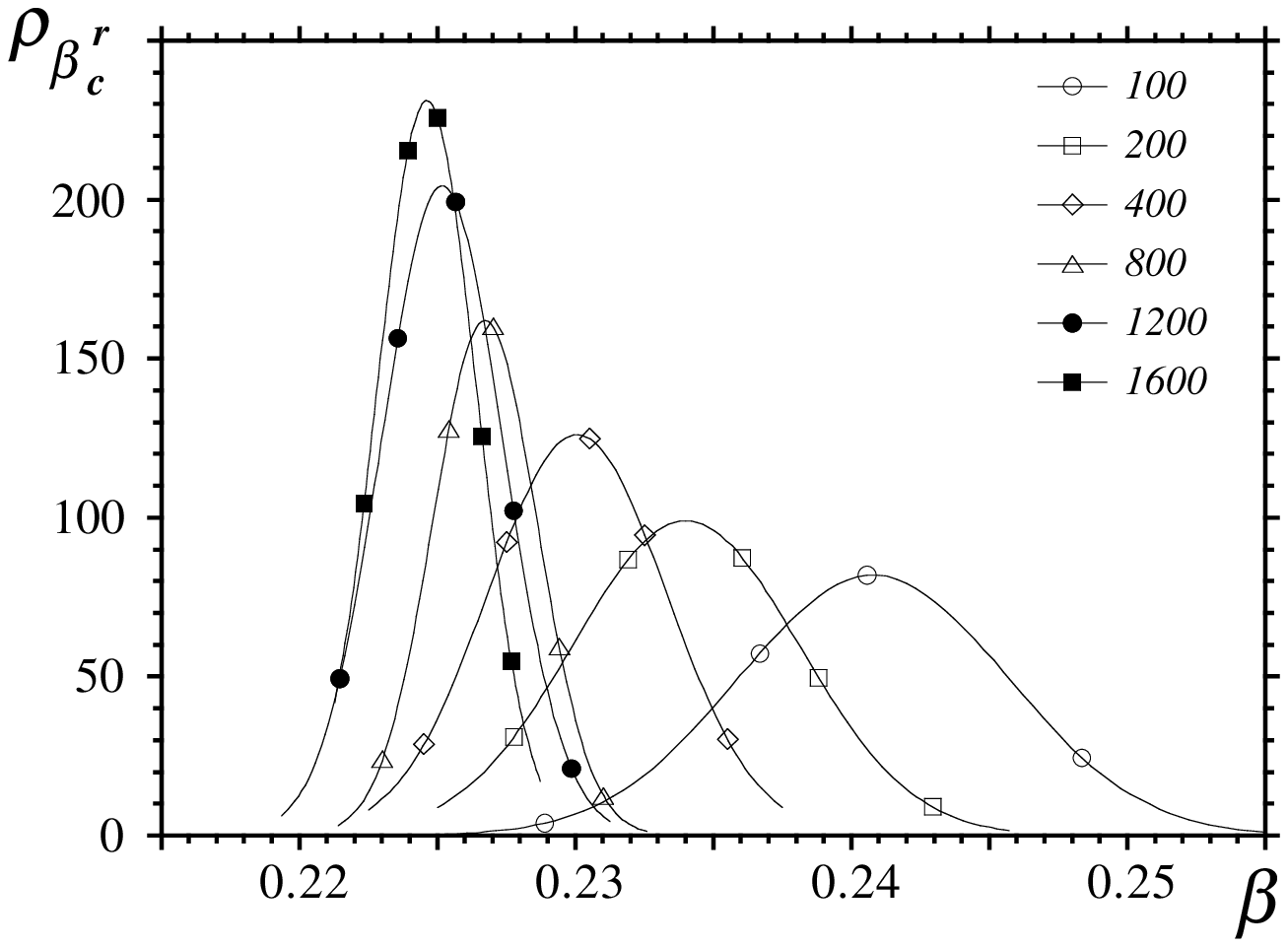}}
\caption{\label{histcv}\small 
 Fits of the (normalized) distributions of 
 the pseudo-critical temperature $\beta_c^r(A)$ 
 to a Gaussian distribution:
 $\rho(\beta_c^r) = a \;
  {\rm exp}(-(\beta_c^r -\bar{\beta}_c)^2/b)$.
  This is for $\beta_c^r(A)$ defined by peaks in
  the specific heat $C^r_A$, for an Ising model simulated
  on different replicas $r$ of $c=-5$ graphs.
  Note that the symbols are only to identify distributions
  corresponding to different area.}
\end{figure}

\begin{figure}
\epsfxsize=4in \centerline{ \epsfbox{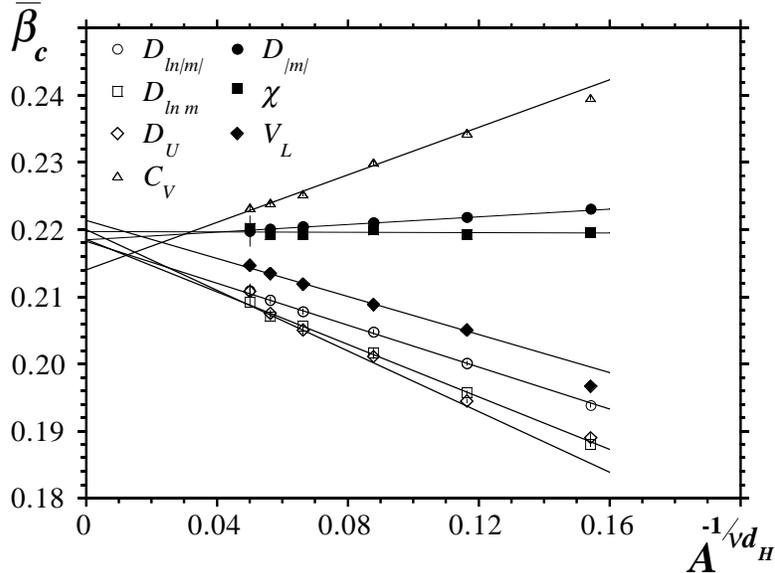}}
\caption{\label{bcscal}\small Scaling of the pseudo-critical temperature 
 $\bar{\beta}_c(A)$  {\it vs}.\ the scaled area $A^{-1/\nu d_H}$, 
 using $\nu d_H = 2.464$.  This is shown for all the different observables
 we considered.  The approach to the (infinite-area) critical 
 temperature $\bar{\beta}_c$ is demonstrated by a linear fit to 
 Eq.~(\ref{coupscal}) (excluding the smallest system size).}
\end{figure}

The fit to Eq.~(\ref{coupscal}) is made considerably easier
by an independent determination of the critical exponent
$\nu d_H$.  The peak values of the derivatives
$\bar{\rm D}_{\ln |m|}$, $\bar{\rm D}_{\ln m^2}$, 
and $\bar{\rm D}_{U_A}$ all scale like $A^{1/\nu d_H}$.
We have used this to determine $\nu d_H$ by a linear
fit; those exponents are shown in Table~\ref{exponents} and 
the corresponding fits in Figure~\ref{bcscal}.  To demonstrate the
finite-size effects we have done the fits with both 
$A_m = 100$ and $A_m = 200$ as the smallest area included in
the fit.  Although the exponents are not substantially
altered by excluding graphs of area 100, for most observables
the $\chi^2$-value of the fit is, however, unacceptably large 
if they are included. 

\setlength{\tabcolsep}{12pt}
\begin{table}
 \begin{center}
 \caption{\label{tabbc}\small The critical temperature 
  $\bar{\beta}_c$, for the Ising model on $c=-5$ graphs, 
  determined from the
  scaling of the location of peaks in the different observables.
  This is both for  $A_m = 100$ and 200 as the smallest
  area include in the linear fit to Eq.~(\ref{coupscal}).
  Also included is the $\chi^2$-value (per {\it d.o.f.}) for
  the fits.}
 \vspace{8pt}
 {\small 
 \begin{tabular}{|l|lc|lc|} \hline
 \vspace{-10pt} & & & &  \\ 
 & \multicolumn{2}{|c}{$A \in\{100,1600\}$}
 & \multicolumn{2}{|c|}{$A \in\{200,1600\}$}   \\
 $\bar{\cal O}$  & $\bar{\beta}_c$ & $\chi^2$ 
                 & $\bar{\beta}_c$ & $\chi^2$ \\ 
 \vspace{-10pt} & & & &  \\  \hline
 \vspace{-10pt} & & & &  \\ 
 $\bar{C}_A$          &  0.2153(15) &  19   &  0.2140(17) &  7.9  \\
 $\bar{\chi}$         &  0.2197(11) &  6.7  &  0.2197(16) &  7.9  \\ 
 $\bar{U}_A$          &  0.2233(12) &  5.6  &  0.2214(10) &  2.1  \\
 $\bar{D}_{\ln |m|}$  &  0.2187(5)  &  1.3  &  0.2183(7)  &  0.8  \\
 $\bar{D}_{\ln m^2}$  &  0.2188(7)  &  4.5  &  0.2186(10) &  4.0  \\
 $\bar{D}_{|m|}$      &  0.2184(5)  &  0.3  &  0.2185(4)  &  0.1  \\
 $\bar{D}_{U_N}$      &  0.2188(22) &  15   &  0.2200(13) &  2.0  \\
 \vspace{-10pt} & & & &  \\  \hline
 \vspace{-10pt} & & & &  \\  
 Average        &  0.2188(9)  &       &  0.2187(9)  &   \\\hline
 \end{tabular} }
 \end{center}
\end{table}

Using the average value $\nu d_H = 2.464(19)$, we obtain the 
critical temperature from a linear fit to Eq.~(\ref{coupscal}).  
The result, for the different observables we considered, 
is shown in Table~\ref{tabbc}. 
This yields the average value $\bar{\beta}_c = 0.2187(9)$
(using $A_m = 200$).

\subsection{The critical exponents}

We then proceed to determine the critical exponents of the
model by fitting the different observables to their
expected finite-size scaling behavior, Eqs.~(\ref{eqobs1}) and 
(\ref{der}), both at the critical temperature and, were appropriate,
also their peak values.  The exponents,
obtained from those fits, are collected in 
Table~\ref{exponents}. 

\setlength{\tabcolsep}{5pt}
\begin{table}
 \begin{center}
 \caption{\label{exponents} \small 
  The critical exponents of the Ising model on a
  quenched ensemble of $c=-5$ graphs, determined from the scaling
  of various observables both at the critical temperature
  $\bar{\beta}_c \approx 0.2187$ and, where appropriate, of peak 
  values.  Graphs of area $A \geq A_m$, with $A_m = 200$
  and 400, are included in the fits to Eqs.~(\ref{eqobs1}) and
  (\ref{der}) (except for $\bar{C}_A$ where it was not possible to 
  obtained a reliable fit if area 200 was excluded). }
 \vspace{8pt}
 {\small
 \begin{tabular}{|ll|lc|lc|lc|lc|} \hline
  \rule[-2mm]{0pt}{7mm}  & & \multicolumn{4}{|c}{\it (A) Scaling of peaks}
      &\multicolumn{4}{|c|}{\it (B) Scaling at $\bar{\beta}_c$} 
      \\\cline{3-10}
 \rule[-2mm]{0pt}{7mm}  & $\bar{\cal O}$  
    & $A_m=200$ & $\chi^2$ & $A_m=400$ & $\chi^2$
    & $A_m=200$ & $\chi^2$ & $A_m=400$ & $\chi^2$ \\\hline
                 & & & & & & & & &\\
   $\nu d_H$  &  $\bar{D}_{\ln |m|}$
  &  2.414(20)  &  7.8  &  2.469(37)  &  2.3
  &  2.413(15)  &  1.0  &  2.451(28)  &  0.4 \\
             &  $\bar{D}_{\ln m^2}$
  &  2.405(23)  &  9.4  &  2.456(23)  &  1.3
  &  2.399(22)  &  2.4  &  2.431(35)  &  0.9 \\
             &  $\bar{D}_{U_N}$
  &  2.591(43)  &  2.4  &  2.528(72)  &  1.7
  &  2.50(12)   &  4.3  &  2.45(11)   &  3.3 \\
                 & & & & & & & & & \\ 
  $\beta/\nu d_H$  & $|m|$
  & & & &
  & 0.1055(34)  &  3.0  &  0.1006(67) &  0.5 \\
           & $\bar{D}_{|m|}$
  & 0.1001(69)  &  1.3  &  0.1070(58) &  0.1
  & 0.0967(52)  &  2.5  &  0.1064(79) &  1.0 \\
                 & & & & & & & & &\\
  $\gamma/\nu d_H$  & $\bar{\chi}$
  & 0.7966(31)  &  2.9  & 0.7922(99)  &  2.6 
  & 0.7825(57)  &  6.8  & 0.7958(146) &  4.0\\
                 & & & & & & & & & \\
  $\alpha/\nu d_H$  & $\bar{C}_A$
  & -0.266(90)  &  0.7  &  &
  & -0.269(78)  &  0.4  &  &  \\\hline
 \end{tabular} }
 \end{center}
\end{table}

The exponents have been determined both for
$A_m = 200$ and 400, except for
the specific heat which is the most difficult quantity to
analyze.  It requires a non-linear 3-parameter fit to the 
scaling behavior Eq.~(\ref{eqobs1}) and, in addition, as
it does not diverge for this particular model
the regular background term is all the more important.  
Thus we could not get a stable fit if we excluded
graphs smaller that 400 and, as is apparent from
the quoted errors, the estimate of this exponent is the
least reliable.  On the other hand, for the other exponents,
most determined from more that one observable, we get very
consistent estimates.  In all cases graphs of area 100
had to be excluded in order to obtain an acceptable fit.    

Although the triangulations we have employed in this
study are smaller than those usually used in simulations
of two-dimensional gravity --- due to the difficulty
in simulating the non-local action Eq.~(\ref{eq213}) --- this is
to a large extent compensated by a smaller fractal
dimension, which sets the relevant length-scale 
that controls the finite-size effects. Hence we are able 
to obtain reliable estimates of the critical exponents,
even from graphs of such modest size.

We have collected in Table~\ref{comparexp}
our numerical estimates of the exponents of an Ising
model on a quenched ensemble of $c=-5$ graphs. 
The exponents shown are 
a weighted average over the values in Table~\ref{tabbc}
(corresponding to $A_m = 400$).  For comparison, we also
included in the table the critical exponents for the 
Ising model on a flat lattice (the Onsager exponents),
for the Ising model coupled dynamically to gravity ($c=1/2$),
and for the Ising model coupled dynamically to gravity
and conformal matter with central charge $c_M=-11/2$ (or 
$c = -5$).  Comparing those exponents it is clear
that our result agrees very well with the last set of exponents.  
That the critical behavior agrees
so well with the predictions from the KPZ scaling
relation for $c=-5$ matter coupled to gravity, strongly
supports our conjecture about the effect of 
disorder on the critical behavior. That is, the 
dynamical interaction between the matter and
the geometry is not important as such, only that they
result in a well defined (average) fractal structure
for the surfaces.

\begin{table}
 \caption{ \label{comparexp} \small The critical exponents of 
  the Ising model on a flat two-dimensional lattice 
  (the Onsager exponents), and coupled dynamically to 
  $2d$-gravity both for $c=1/2$ and $c=-5$. 
  This is compared to the results of our simulations for
  the Ising model on a quenched ensemble of $c=-5$ graphs.
  Also shown are the corresponding exponents for the 
  fractal structure of the graphs: $\gamma_s$ and $d_H$.}
 \begin{center}
 \vspace{4pt}
 {\small
 \begin{tabular}{|c|cccc|cc|} \hline
    & $\nu d_H$       & $\alpha/\nu d_H$  &  $\beta/\nu d_H$ 
    & $\gamma/\nu d_H$ & $\gamma_s$       &  $d_H$ \\       
   \vspace{-10pt} & & & & & &  \\ \hline
   \vspace{-10pt} & & & & & &  \\ 
   {\it Onsager}  &  2   &  log  &  0.0625  &  0.875  & $-\infty$ & 2  \\
   $c = -5$       &  2.452\ldots & -0.184\ldots   & 0.0955\ldots
                  &  0.809\ldots & -1.618\ldots   & 3.236\ldots    \\ 
   $c = 1/2$      &  3           & -0.333\ldots   & 0.1666\ldots  
                  &  0.666\ldots & -0.333\ldots   & 4.0-4.2         \\
   \vspace{-5pt}  & & & & & &  \\
   {\it Quenched} &  2.464(19)   & -0.27(8)       & 0.105(4)
                  &  0.793(8)    & -1.5(1)        & 3.36(4) \\\hline
 \end{tabular} }
 \end{center}
\end{table}

\section{Discussion}

The main results of the work presented in this paper
can be summarized as follows:
\begin{itemize}
 \item[({\it a})] 
  The fractal dimension of  surfaces, defined by a conformal
  field theory with central charge $c=-5$ coupled to two-dimensional
  quantum gravity, is $d_H = 3.36(4)$.  This is in reasonable
  agreement with, and supports, the theoretical prediction 
  Eq.~(\ref{andH2}), whereas it definitely rules out
  Eq.~(\ref{andH1}).
 \item[({\it b})]
  The critical behavior of an Ising model on a {\it quenched} ensemble
  of $c=-5$ graphs agrees well with the
  predictions, from the KPZ scaling relation, for an Ising
  model on an {\it annealed} ensemble of graphs with {\it identical}
  fractal properties.
\end{itemize}

The first result, especially 
combined with the recent simulations of 
$2d$--gravity for $c=-2$ \cite{cm2}, lends a strong
support to Eq.~(\ref{andH2}) as a correct description
of the fractal structure of two-dimensional quantum 
gravity for $c \leq 0$.  This makes, however, its disagreement with
numerical simulations in the region $0 < c \leq 1$ all the
more surprising.  What is it in derivation of Eq.~(\ref{andH2}) 
that breaks down for $c>0$?
Or are the simulations dominated by finite-size errors 
and simulations of larger systems will eventually agree with
Eq.~(\ref{andH2})? 

The result for the Ising model is even more interesting.  
As the theoretical predictions are obtained for an Ising
model coupled dynamically to the disorder, this supports
the conjecture put forward in the Introduction about the
equivalence between annealed and quenched averages over
disorder. That is, the only
thing relevant for the critical behavior of the Ising model
are the average fractal properties of graphs the spins ``see''. 
How the statistical average over graphs is performed, 
quenched or annealed, is not relevant.

It is also worth noting that we can continuously
change the average fractal properties of the graphs 
by changing the embedding dimension $D$ in Eq.~(\ref{eq213}).
This allows a 
continuous interpolation between a flat surface and 
surfaces corresponding to pure gravity.
If the prediction of Liouville theory, the KPZ formula, holds for all those
models, this implies that the critical behavior of the Ising model
should change continuously in the process.
In the language of
the renormalization group this implies a continuous line
of fixed points, rather than isolated points.  There are 
well known examples of this; the low-temperature phase of
the two-dimensional $XY$--model or the critical line of
the Ashkin-Teller model.
But is this statement also true for very weak disorder?
If we change the fractal dimension infinitesimally,
from 2 to $2+\epsilon$, is that  enough to
change the critical behavior of the Ising model?  Or,
alternatively, does there exist some central charge    
$c^{\prime} < -5$ were the geometrical disorder is not strong 
enough and we always get the Onsager exponents?
This point deserves further study.

One could also look at the examples of weak disorder
that have been studied recently,
for example the site or bond-diluted Ising model,
and ask if that kind of disorder can also be classified 
according to some average fractal properties of the lattices.
And, moreover, if one could observe some kind of
universality in the critical behavior, depending on the
fractal structure, akin to what we have presented 
in this paper.
 
In view of how dramatically the critical behavior of 
the Ising model changes on surfaces with such strong
disorder, one might ask if such change could be observed in
real physical systems. Possible candidates for such
systems could be, for example, electrons trapped on the 
interfaces between two liquids, or on the surface of 
some porous material, were the surfaces had some well
defined non-trivial fractal structure.
As our results indicate, it is only the average geometry
of the surfaces that is important for the Ising model, not
its fluid nature or curvature fluctuations.  Thus the relative
time-scale between the interactions of the particles and the
change in the geometry should be irrelevant.

\vspace{20pt}
\noindent
{\bf Acknowledgments:}
The work of G.T.\ was supported by the Humboldt Foundation. 
The work of P.B.\ was partially supported by KBN grants
2P03 B19609 and 2P03 B04412.

\end{document}